\documentclass[jkps,graphicx,preprint,showpacs,showkeys]{revtex4-1}

\draft 
\usepackage{graphicx}
\begin{document}


\title{The Control of the Oscillation Threshold with Asymmetric Gain in Operational Amplifiers } 



\author{Sungmook \surname{Kang}}
\author{Heeso \surname{Noh}}
\email[]{heesonoh@kookmin.ac.kr}
\thanks{Fax: +82-2-910-4728}
\affiliation{Department of Nano and Electronic Physics, Kookmin University, Seoul 02707, Korea}


\date{\today}

\begin{abstract}
We have observed the quench of the lasing at the exceptional point in the electronic circuit system by applying asymmetric gain in the coupled oscillator. Since there is the analogy between oscillation in laser and oscilation in the operational amplifier, when the system hits the exceptional point, oscillation stops. This phenomenon is also theoretically investigated.   
\end{abstract}

\maketitle 

\section{Introduction}

If a potential $V$ of a physical system satisfies $V({-\bf x})=V^*({\bf x})$ (${\bf x}$ is position vector and $^*$ means complex conjugate), we call such potential parity time ($PT$) symmetric. The system with $PT$ symmetric potential has been attracting a lot of interest because it can have real valued eigenvalues even though the potential of the system is not Hermitian.\cite{bender1998,bender2007} The more interesting part of the $PT$ symmetric potential is that it has "threshold". Below the threshold, the eigenvalues are real (exact phase) while the eigenvalues are imaginary (broken phase) above the threshold. 
Since $PT$ symmetric system was first reported, it has been also applied to optics by the use of special complex refractive index $n$ satisfying the relation $n({-\bf x})=n^*({\bf x})$. Researches such as lasing in $PT$ symmetric potential, one way transmission etc. have been reported.\cite{guo2009,Ruter2010,Regensburger2012,Longhi2009,lin2011,Chong2011,Feng2014}

At threshold, $PT$ symmetric system goes into a special state, called exceptional point. At the exceptional point, eigenmodes are coalesced.\cite{Heiss2012,Ramezani2012} the exceptional point behavior has been investigated in the $PT$ symmetric system. Liertzer, M. et al.\cite{Liertzer2012} proposed the suppression of lasing oscillation due to exceptional point. This idea was realized in the optical system using two micro-disk lasers and electrical system using two $RLC$ resonators.\cite{Chitsazi2014,Brandstetter2014} In electrical system, it is better to show the suppression of lasing oscillation using active electric component because the operational amplifier (op-amp) is analog to the laser such that  the gain of the op-amp can be controlled with resistors and the output signal of the op-amp by employing positive feedback have oscillation threshold.              

In this Letter, we demonstrate the suppression of oscillation using two operational ampplifiers (op-amps). By adjusting the gain of the op-amps, the output signal from the op-amp stops oscillation at the exceptional point.
\section{Experiment}

\begin{figure}[htbp]
\includegraphics[width=4in]{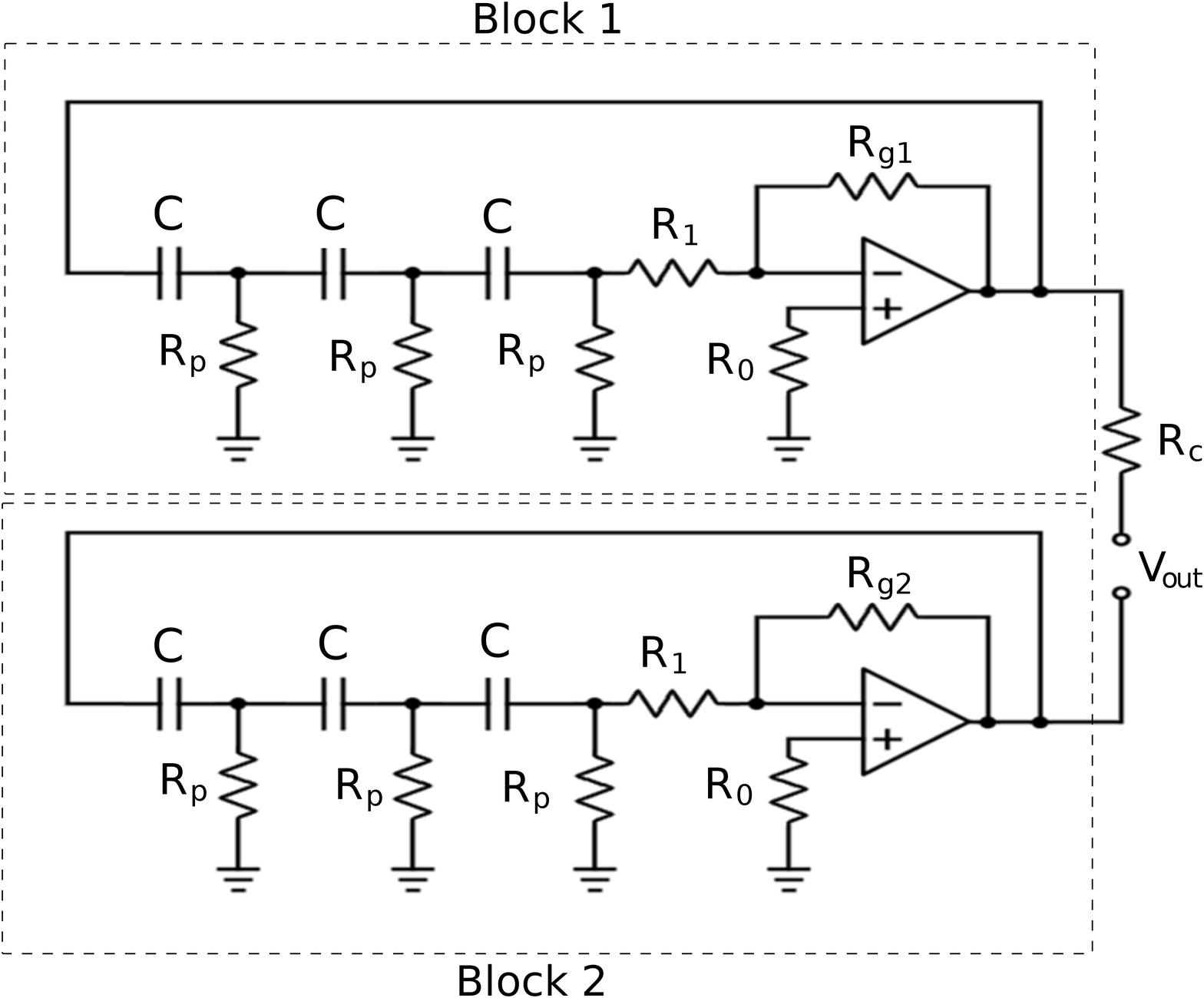}
\caption{circuit diagram. Two op-amps with positive feedback are connected with a resistor $R_c$.}
\label{fig1}
\end{figure}

Figure \ref{fig1} shows the circuit diagram. Block 1 and block 2 contain identical circuits, and two blocks are connected through a resistor $R_c$ (168.9 $k\Omega$) which adjusts the coupling between block 1 and block 2. The op-amps we use are OP-27's. The amount of amplification of the op-amp is determined by $R_{g1}/R_1$ for block 1 and $R_{g2}/R_1$ for block 2. $R_1$ is set to 10 $ k\Omega$. $R_{g1}$ and $R_{g2}$ are variable resistors since we need to change gain in each block separately. Three sets of $R_p$-$C$ pairs in each block give the phase shift. The resistance of $R_p$ is  2 $k\Omega$, and the capacitance of $C$ is 10 $nF$. Oscillation frequency of output signal from each block is determined by the amount of phase shift. The output frequency is the frequency where the phase shift becomes $\pi$ (positive feedback). We also put $R_0= 10\;k\Omega$ for stable operation.

\section{Results}

\begin{figure}[htbp]
\includegraphics[width=3.5in]{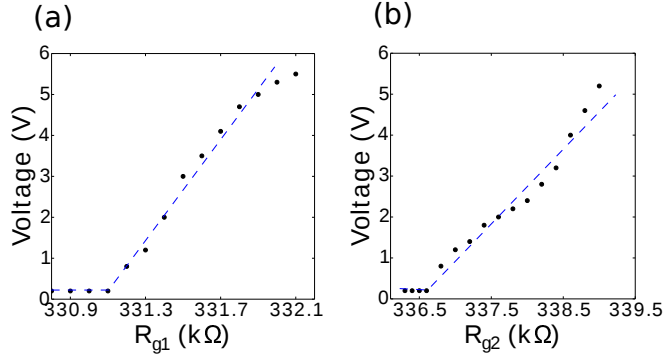}
\caption{(a) output voltage of op-amp vs. $R_{g1}$in block 1.(b) output voltage of op-amp vs. $R_{g2}$ in block 2. By changing $R_{g1}$ and $R_{g2}$, the gain of the op-amps can be determined. The kinks in (a) and (b) show oscillation thresholds. The resistance of $R_{g1}$ at the oscillation threshold is  331.1 $k\Omega$ and the resistance of $R_{g2}$ is 336.7 $k\Omega$. }
\label{fig2}
\end{figure}

We measured the oscillation threshold of the output voltage from block 1 (block 2) by changing the gain of the op-amp using variable resistor $R_{g1}$ ($R_{g2}$) with an oscilloscope. The peak value of the output voltage is depicted in Fig. \ref{fig2}. When Resistance $R_{g1}$ is 331.1 $k\Omega$, the slope of output voltage changes abruptly (Fig. \ref{fig2}a). This kink indicates the oscillation threshold. This is similar to the laser because the laser starts to lasing at the threshold and the threshold can be determined by the slope change of the output intensity of the laser when pumping power increases. The resistance $R_{g2}$ measured at oscillation threshold in block 2 is 336.7 $k\Omega$ (Fig. \ref{fig2}b). One can see almost the same resistances of $R_{g1}$ and $R_{g2}$ at oscillation threshold. The slight difference is attributed to the tolerance of resisters which is $5 \%$.

\begin{figure}[htbp]
\includegraphics[width=3.5in]{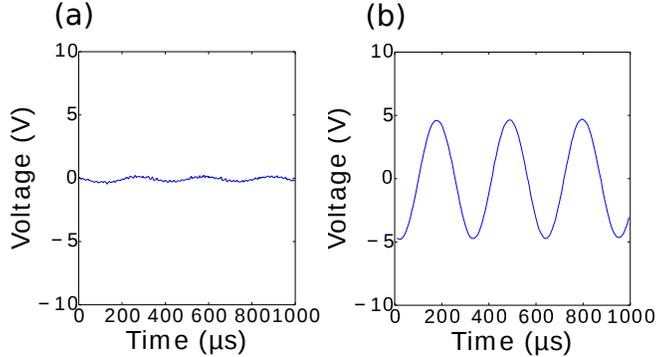}
\caption{Output voltage vs. time from the op-amp in block 1 below threshold (a) and above threshold (b). One can see the oscillation of output voltage above the oscillation threshold.}
\label{fig3}
\end{figure}

Figure \ref{fig3} shows output voltage of block 1 below and above oscillation threshold. Below the oscillation threshold, Op-amp does not oscillate (Fig. \ref{fig3} (a)). However above the oscillation threshold, Oscillation started (Fig. \ref{fig3} (b)). 

The oscillation frequency can be estimated by the following equation
\begin{equation}
 f=\frac{1}{2\pi CR\sqrt{6}}.
\end{equation}
 Since the capacitance $C$ is $10 nF$ and the resistance $R$ is $2 k\Omega$, the estimated oscillation frequency is 3.25 kHz. The measured frequencies are 3.44 kHz for block 1 and 3.38 kHz for block 2. The frequencies of two oscillators from two blocks and estimated frequency are in good agreement. The difference between the estimated and measured frequencies is only about $5\; \%$. This difference is mainly attributed to the tolerance of resistance $R$ which is $5\;\%$. Therefore slight difference of oscillation frequencies can be understood if the tolerance of the resistor is taken into account. Those oscillation frequencies do not change by changing gain above the oscillation threshold. 

Two oscillators are connected via resistor $R_c$. By adjusting the value of $R_c$, we can control the coupling between two oscillators block 1 and block 2. As the value of $R_c$ increases (decreases), the coupling strength between two oscillators becomes weaker(stronger). We set $R_c=168.9\Omega$ in the experiment.

Since the threshold resistance of the oscillator in block 1 is $331.1 k\Omega$, we fixed the resistance at $338k\Omega$ which is above the threshold. And since the threshold resistance of block 2 $336.7 k\Omega$, we change the resistance from $334k\Omega$(below threshold) to $338k\Omega$ (above threshold). 

Figure \ref{fig4}(a) shows the change of the oscillation amplitude as resistance $R_{g2}$ increases. Initially oscillation amplitude $V_{out}$ is 5 V. As the $R_{g2}$ increases, $V_{out}$ decreases. When $R_{g2}$ becomes 336 $k\Omega$ oscillation stops. If we increase the resistance $R_{g2}$ further, $V_{out}$ increases again. and the oscillation amplitude is recovered to 5 V. The even further increase of the resistance $R_{g2}$ makes the oscillation amplitude larger. This result is counter intuitive because if there is no coupling between block 1 and block 2, the amplitude of output oscillation would increase monotonically. 

\section{Discussion} 

It can be interpreted as follows. If we assume two coupled cavities, The Hamiltonian for the system is,  
\begin{equation}
H=\left(\begin{array}{c c}\alpha & \gamma \\ \gamma & \beta
\end{array}\right)
\label{eq2}
\end{equation}
where $\alpha$ and $\beta$ are complex numbers. Real parts (Re) of them are related resonance frequencies. Imaginary parts (Im) of them indicate absorption or amplification. For example, positive number of Im($\alpha$) indicates amplification (lasing),negative number absorption, and zero means oscillation threshold. $\gamma$ is coupling constant. Therefore this matrix is a simple description of two oscillator system with coupling. 

\begin{figure}[htbp]
\includegraphics[width=3.5in]{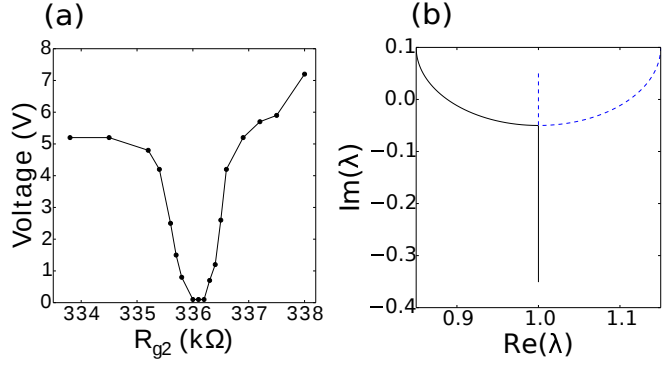}

\caption{(a) experimental results: when we incresese gain ($R_{g2}$), initially $V_{out}$ decreases and stop oscillation, However when we increase further the oscillation is recovered. (b)The change of two eigenvalues when we increases gain ($\beta$). Blue dashed curve is for $\lambda_1$ and black solid curve is for $\lambda_2$. As we increase $\beta$ from -0.4 to 0.1, $\lambda_1$ changes from $1.0+0.35i$ to $1.2+0.1i$ and $\lambda_2$ changes from $1.0-0.35i$ to $0.8+0.1i$.}
\label{fig4}
\end{figure} 

We investigate the the change of the eigenvalues as Im($\beta$) increases in the Hamiltonian (eq. \ref{eq2}). In order to simulate the experiment we performed, we set $\alpha = 1+0.1i$ which is above threshold,$\gamma=0.15$ for coupling of two cavities. We also set Re($\alpha$)=1 because oscillation frequencies of two cavities are the same. After that we increased Im($\beta$) from -0.4 to 0.1 which means that gain is increased. Since matrix is 2 by 2 matrix, generally two eigenvalues can be obtained. We plot two eigenvalues in Fig. \ref{fig4} (b). Blue dashed curve is for one eigenvalue ($\lambda_1$), and black solid curve is for the other eigenvalue ($\lambda_2$). When we change $\beta$ from -0.4 to -0.2, the real parts of $\lambda_1$ and $\lambda_2$ have the same values as 1. Only imaginary parts of them change. Im($\lambda_1$) decreases from 0.05 to -0.05, and Im($\lambda_2$) increases from -0.35 to -0.05. At $\beta$ = -0.2, two eigenvalues become the same value as $1-0.05i$. This is exceptional point where two eigenmodes are coalesced. Therefore when we change $\beta$ from -0.4 to -0.2, two eigenvalue move toward the exceptional point. As $\beta$ changes from -0.2 to 0.1, real parts of $\lambda_1$ and $\lambda_2$ move away from the exceptional point in opposite direction and imaginary parts of $\lambda_1$ and $\lambda_2$ move away from the exceptional point but they are the same value.

This simulation result can be compared to the experimental result. When $\beta$ is between $-0.4$ and $-0.225$, Im($\lambda_1$) decreases but still it is positive number while Im($\lambda_1$) is negative. This means that the eigenmode with $\lambda_1$ is above the oscillation threshold while the eigenmode with $\lambda_2$ is below the oscillation threshold. Since one eigenvalue is above the oscillation threshold, one can see the strong oscillation. When $\beta$ is between -0.225 and -0.1, both eigenvalues are below oscillation threshold. Therefore strong oscillation stops. As we keep increase the imaginary parts of both eigenvalues have the same positive number. Strong oscillation starts again. This is similar to the experimental results. Initially $V_{out}$ shows strong oscillation. As we increase the $R_{g2}$, oscillation stops. When we increase $R_{g2}$ further oscillation starts again. This means that exceptional point is responsible for the quench of oscillation.

\section{Conclusion}

We made a circuit consisting of two identical op-amps with positive feedbacks, where two op-amps are coupled. When we keep one op-amp above the oscillation threshold, and increase the gain of the other op-amps from the below threshold to the above threshold, we observe the quench of the output oscillation. This quench of oscillation is due to the exception point according to numerical simulation. Therefore we show that by using exceptional point, we can control the osillation threshold. Our results can be applied to the modulation of output signal in an active system.  



%
%

%

\begin{acknowledgments}
This work was supported by the Basic Science Research Program through the National Research Foundation of Korea (NRF) funded by the Ministry of Education, Science, and Technology (No.2013R1A1A1010968).
\end{acknowledgments}

\bibliography{EP.bib}

\end{document}